\documentclass[11pt,a4paper]{article}

\usepackage{jheppub}
\usepackage{bm}

\title{Renormalized vs Nonrenormalized Chiral Transition in a Magnetic Background}

\author[a]{Marco Ruggieri,}
\author[b]{Motoi Tachibana,}
\author[a,c]{and Vincenzo Greco}
\affiliation[a]{Department of Physics and Astronomy, University of Catania, Via S. Sofia 64, I-95125 Catania, Italy.}
\affiliation[b]{Department of Physics, Saga University, Saga 840-8502, Japan.}
\affiliation[c]{INFN-Laboratori Nazionali del Sud, Via S. Sofia 62, I-95123 Catania, Italy.}

\emailAdd{marco.ruggieri@lns.infn.it}
\emailAdd{motoi@cc.saga-u.ac.jp}
\emailAdd{greco@lns.infn.it}

\abstract{
We study analytically the chiral phase transition for hot quark matter in presence of
a strong magnetic background, focusing on the existence of a  critical point 
at zero baryon chemical potential and nonzero magnetic field.  
We build up a Ginzburg-Landau
effective potential for the chiral condensate at finite temperature, computing 
the coefficients of the expansion within a chiral quark-meson model.  
Our conclusion is that the existence of the critical point
at finite $\bm B$ is very sensitive to the way the ultraviolet divergences 
of the model are treated. In particular, we find that after renormalization, 
no chiral critical point is present in the phase diagram.
On the other hand, such a critical point there exists when the 
ultraviolet divergences are not removed by a proper renormalization
of the thermodynamic potential. 
}

\keywords{Chiral phase transition in a magnetic background.}

\begin{document}
\maketitle
\flushbottom

\section{Introduction}
Recent simulations of relativistic heavy ion collisions suggested the intriguing possibility that
huge magnetic fields are created during noncentral collisions~\cite{Kharzeev:2007jp,Skokov:2009qp,Voronyuk:2011jd}. 
At the typical Relativistic Heavy Ion Collider energies
the current estimate for the largest magnetic field produced is $eB \approx 5 m_\pi^2$, 
with $m_\pi$ corresponding to the pion mass in the vacuum\footnote{It is customary
to measure the value of $eB$ in units of the vacuum pion mass. 
As a rule of thumb, to $eB = m_\pi^2$ corresponds
$B \approx 10^{14}$ Tesla.}.
Moreover, at typical Large Hadron Collider energies it is found $eB \approx 15 m_\pi^2$ as a maximum value for the 
produced magnetic field. 
Electric fields are also produced in the collisions, but their magnitude is smaller than that of their magnetic counterpart.
Simulations show that the structure of the electromagnetic fields in space and time is rather complicated, and in fact
these fields are highly inhomogeneous and short lived. On the other hand, these results make the study of
the role of the electromagnetic fields on the phase structure of Quantum Chromodynamics (QCD) not academic.
Relevant studies can be found in the book~\cite{Kharzeev:2012ph} as well as in
\cite{Bali:2011qj,Klevansky:1989vi,D'Elia:2011zu,Suganuma:1990nn,Gusynin:1995nb,Frasca:2011zn,Klimenko:1990rh,Fraga:2008qn,
Agasian:2008tb,Bruckmann:2013oba,D'Elia:2010nq,Endrodi:2013cs,Fukushima:2010fe,Mizher:2010zb,Andersen:2012bq,
Skokov:2011ib,Fukushima:2012xw,Chernodub:2010qx,Buividovich:2008wf,Hidaka:2012mz,Fraga:2012fs,Preis:2010cq,
Callebaut:2013ria,Andersen:2011ip,Burikham:2011ng,Filev:2010pm,Ferrari:2012yw,Gynther:2010ed,Mizher:2013mia}.
The existence of strong fields in heavy ion collisions, combined to the excitation of strong
sphalerons at high temperature~\cite{Moore:1997im}, also suggested the possibility of event-by-event
${\cal P}-$ and ${\cal CP}-$odd effect dubbed Chiral Magnetic Effect~\cite{Kharzeev:2007jp,Fukushima:2008xe,
Rebhan:2009vc,Kharzeev:2010gr,Gorsky:2010xu,Braguta:2010ej,Gahramanov:2012wz,Nam:2009jb,Wang:2012qs,Voloshin:2010ut},
see~\cite{Fukushima:2012vr,Kharzeev:2012ph} for reviews.  
Besides heavy ion collisions, even stronger magnetic fields might have been produced in the early universe
at the epoch of the electroweak phase transition, $t_{ew}$, because of gradients in the vacuum expectation
value of the Higgs field: a widely accepted value for the magnetic field at the transition
is $B(t_{ew}) \approx 10^5 m_\pi^2\approx 10^{19}$ T, even if this value rapidly decreased 
scaling as $a^{-2}$, where $a(t)$ denotes the scale factor of the expanding universe,
losing several order of magnitude at the QCD phase transition.   
Finally, strong magnetic fields (but still weaker than the one produced
in relativistic heavy ion collisions) are present on the surface of magnetars,
$B \approx 10^{-4} m_\pi^2\approx 10^{10}$ T~\cite{Duncan:1992hi}.
Therefore, there exist at least three physical contexts in which 
QCD in a strong magnetic background is worth to be studied.

There is general consensus on the fact that a strong magnetic field affects the physics
of spontaneous chiral symmetry breaking in QCD. At zero temperature, the magnetic field
induces the magnetic catalysis, leading to an increase of the chiral condensate with the strength
of the applied field.  
On the other hand, the effect of the external field on the restoration
of chiral symmetry at finite temperature is still controversial. In fact recent lattice
simulations have shown that increasing the strength of the magnetic field leads to a reduction
of the critical temperature for chiral restoration~\cite{Bali:2011qj}, 
which is understood in terms
of the backreaction of the quark loops on the gluon action~\cite{Bruckmann:2013oba}.
This conclusion is in agreement with some
previous model computations, see for example~\cite{Agasian:2008tb} 
and~\cite{Mizher:2010zb},
but in disagreement with most of the recent chiral model computations.
The lattice results show as well some role of the bare quark mass on the shift
of the critical temperature; on the other hand, model computations have not been able
to describe such a behavior, see for example~\cite{Mizher:2013mia}. 
Interesting possibilities to solve this puzzle can be found in~\cite{Kojo:2012js}.

Leaving apart the problem of the critical temperature, there is general agreement that
the magnetic field makes the chiral phase transition stronger, and it might turn to 
a first order phase transition at large enough $\bm B$. 
If this is the case, then a critical point in the 
$(T,\bm B)$ plane should appear at zero baryon chemical potential~\footnote{In~\cite{Agasian:2008tb} 
such critical endpoint is found; however, it connects a first order transition at small $\bm B$
with a second order transition at large $\bm B$.}.
Lattice results are in agreement with the strengthening of the phase transition,
but they show no evidence for a critical point. The existence of the latter
is also not confirmed by models: some model computations predict a first order
phase transition if $eB$ is large enough, while other models do not observe
a change of the phase transition from crossover to first order even for large 
magnetic field strengths.

In this Article we address analytically the problem of the chiral phase transition 
for quark matter at zero baryon chemical potential and nonzero magnetic field,
focusing on the possible existence of a critical point in the phase diagram and
connecting it to the ultraviolet divergences of the underlying microscopic model.  
In order to make quantitative predictions, we build up a Ginzburg-Landau (GL)
effective potential for the chiral condensate at finite temperature; neglecting
inhomogeneous condensates as well as terms which are not dependent on the condensate, we write the
effective GL action around the critical line as
\begin{equation}
\Omega = \frac{\alpha_2}{2}m_q^2 + \frac{\alpha_4}{4}m_q^4~,
\label{eq:i1}
\end{equation}
where $m_q$ corresponds to the dynamical quark mass which is proportional to the chiral condensate,
see the description of the model below.
According to the general GL description of a phase transition, the order of the latter depends on the sign
of the coefficient $\alpha_4$ at the temperature $T_c^*$ defined as the solution of the equation $\alpha_2(T^*_c) = 0$.
If $\alpha_4(T^*_c) > 0$ then the phase transition is of the second order and the critical temperature is $T_c = T^*_c$; 
on the other hand, if $\alpha_4 (T_c) < 0$ then the phase transition is of the first order and the critical temperature
satisfies $T_c > T^*_c$. Hence the point $(\alpha_2,\alpha_4)=(0,0)$ in the phase diagram corresponds to 
a critical point, at which a first order and a second order phase transition lines meet. 

For the mapping of the phase diagram from the $(\alpha_2,\alpha_4)$ plane to the
$(T,\bm B)$ plane we need a microscopic model to compute the explicit
dependence of the GL coefficients on these variables. In this Article we make use
of the chiral quark-meson model~\cite{Jungnickel:1995fp,Herbst:2010rf,Schaefer:2004en,Skokov:2010wb}, 
and restrict ourselves to the one-loop approximation
which amounts to neglect quantum fluctuations for the meson fields in the
thermodynamic potential. The advantage of the quark-meson model is its renormalizability,
which allows to make quantitative predictions which are not affected by any ultraviolet scale.
In fact, our main goal is to elucidate on the role of the ultraviolet divergences in this model
on the order of the phase transition.
Our main conclusion is that a critical point
in the $(T,\bm B)$ plane might exist, but its existence is very sensitive to the way the ultraviolet divergences 
of the model are treated. In particular, we are able to predict analytically that after 
the renormalization of the thermodynamic potential has been properly performed,  
no chiral critical point exists.
On the other hand, such a critical point there exists when the 
ultraviolet divergences are not removed by a proper renormalization. 

We stress that even if the model parameters are fixed having in mind 
the QCD spontaneous chiral symmetry breaking, our results are
general and apply to any theory of charged interacting fermions
in a magnetic background, when a scalar condensate appears in the theory,
which we dub QCD-like theories according to a widely spread nomenclature.
Keeping this in mind, it is not the scope of this Article to investigate 
on the discrepancy between model and Lattice computations about the effect of the magnetic
field on the critical temperature of the chiral phase transition.
Adding by hand a term that leads to a decreasing critical temperature in the GL
second order coefficient would be an easy exercise; 
however we do not follow this procedure since we are not interested to   
a quantitative comparison with the Lattice simulations, but to
a quantitative study about how  ultraviolet divergences
in QCD-like theories affect the physical predictions.
Moreover, adding this $\bm B-$dependent term in the GL effective
action naturally requires a further arbitrary term in the
quartic coefficient, thus ruining the quantitative power of our study
and the possibility to compare directly our results to those obtained within
QCD-like theories. 

We also stress that our purpose is not to claim that previous model computations lacking renormalization
are wrong. In fact, the explicit 
ultraviolet cutoff appearing in the model calculations is a signal of a rough modelling
of the QCD asymptotic freedom: the interactions are switched off 
for momenta larger than the ultraviolet scale. Therefore we consider our study
as a tasteful theoretical investigation of the role of the ultraviolet divergences
in the microscopic model on the order of the phase transition, rather than as a censor 
which judges which is the more appropriate procedure to treat the divergences of the model
when it is applied to describe QCD. Nevertheless, beside the interest from the pure
field theory point of view, the results can be nicely interpreted
physically observing that the critical point appears naturally in the model
with the explicit UV cutoff, which is a remnant of the QCD asymptotic freedom,
hence connecting the latter property of the strong interactions to the existence
of the former.

The plan of the Article is as follows: in Section II we describe the renormalization
of the chiral quark-meson model in a magnetic background, discussing how the divergences
can be embedded in the vacuum and $\bm B-$independent thermodynamic potential.  
In Section III we compute the GL coefficients of the thermodynamic potential
in Eq.~\eqref{eq:i1} using the renormalized model. In Section IV we repeat the
same computation using the nonrenormalized model, thus leaving the explicit dependence
on the ultraviolet cutoff. In Section IV we summarize our results and draw our conclusions.

\section{Renormalization of the quark-meson model in a magnetic background}
In this Section, we specify the model we use in our calculations,
explaining its lagrangian density and how we renormalize the vacuum 
part of the thermodynamic potential. 
The lagrangian density of the model is given by
\begin{eqnarray}
{\cal L} &=& \bar q \left[iD_\mu\gamma^\mu - g(\sigma +
i\gamma_5\bm\tau\cdot\bm\pi)\right] q
\nonumber \\
&& + \frac{1}{2}\left(\partial_\mu\sigma\right)^2 +
\frac{1}{2}\left(\partial_\mu\bm\pi\right)^2 - U(\sigma,\bm\pi)~.
\label{eq:LD1}
\end{eqnarray}
In the above equation, $q$ corresponds to a quark field in the
fundamental representation of color group $SU(N_c)$ and flavor group
$SU(2)$; the covariant derivative, $D_\mu = \partial_\mu - Q_f e
A_\mu$, describes the coupling to the background magnetic field,
where $Q_f$ denotes the charge of the flavor $f$. Besides,
$\sigma$, $\bm\pi$ correspond to the scalar singlet and the
pseudo-scalar iso-triplet fields, respectively. The potential $U$
describes tree-level interactions among the meson fields. In this
article, we take its analytic form as
\begin{equation}
U(\sigma,\bm\pi) = \frac{\lambda}{4}\left(\sigma^2
+\bm\pi^2-v^2\right)^2~, \label{eq:L2}
\end{equation}
which is invariant under chiral transformations. 

In this article, we restrict ourselves to the one-loop large-$N_c$ approximation, 
which amounts to consider mesons as
classical fields, and integrate only over fermions in the
generating functional of the theory to obtain the thermodynamic 
potential. As a matter of fact, quantum
corrections arising from meson bubbles are suppressed of a factor
$1/N_c$ with respect to case of the fermion bubble. In the
integration process, the meson fields are fixed to their classical
expectation value, $\langle\bm\pi\rangle = 0$ and
$\langle\sigma\rangle \neq 0$. 
The physical value of $\langle\sigma\rangle$ will be
then determined by minimization of the thermodynamic potential.
This implies that one replaces $g\sigma \rightarrow
g\langle\sigma\rangle$ in the quark action.
The field $\sigma$ carries the quantum numbers of the quark chiral condensate,
$\langle\bar q q\rangle$; hence, in the phase with $\langle\sigma\rangle\neq 0$,
chiral symmetry is spontaneously broken.

The one-loop fermion bubble associated to the interaction with
a magnetic background can be computed within the Leung-Ritus-Wang method~\cite{Ritus:1972ky},
namely
\begin{equation}
\Omega_B = -N_c\sum_f \frac{|Q_f eB|}{2\pi}\sum_{n=0}^\infty\beta_n \int_{-\infty}^{+\infty}
\frac{dp_z}{2\pi}\left[E + 2T\log\left(1+e^{-\beta E}\right)\right]~,
\label{eq:1}
\end{equation}
where $n$ labels the Landau level, $E$ corresponds to the single particle excitation spectrum, 
\begin{equation}
E = \sqrt{p_z^2 + 2 |Q_f eB|n + m_q^2}~,
\end{equation}
and $m_q = g\langle\sigma\rangle$ is the constituent quark mass. The factor $\beta_n = 2 - \delta_{n0}$
counts the degeneracy of the $n^{th}$-Landau level. 

The divergence in $\Omega_B$ is contained in the vacuum contribution. 
Since the model is renormalizable, we can treat this divergence by means of renormalization.
In order to prepare $\Omega_B$ for renormalization, we firstly add and subtract the contribution
at $\bm B = 0$, namely
\begin{equation}
\Omega_{0} = -2N_c N_f \int\frac{d^3p}{(2\pi)^3}\left[\omega 
+ 2T\log\left(1+e^{-\beta\omega}\right)\right]~,
\label{eq:2}
\end{equation}
where $\omega = \sqrt{\bm p^2 + m_q^2}$. This procedure is convenient since it allows to collect
all the contributions due to the magnetic field into an addendum which is ultraviolet 
finite.
In principle it would be enough to consider only the vacuum contribution in $\Omega_0$ (and in fact,
such contribution will be the only one affected by the renormalization procedure); however,
for the computations of the GL coefficients at finite $\bm B$ it is convenient to add and
subtract the $\bm B = 0$ finite temperature contribution as well.
Hence we write
\begin{equation}
\Omega_B = \Omega_0 + \left(\Omega_B - \Omega_0\right) \equiv \Omega_0 + \delta\Omega~.
\label{eq:PP}
\end{equation}  
In the following we renormalize $\Omega_0$; then we discuss the ultraviolet behavior of $\delta\Omega$,
showing that it is finite and hence it does not need to be renormalized.

\subsection{Renormalization of the zero field contribution}
We split the zero field potential into a vacuum and
a valence part, $\Omega_0 = \Omega_0^0 + \Omega_0^T$,
with
\begin{eqnarray}
\Omega_0^0 &=&-2N_c N_f \int\frac{d^3p}{(2\pi)^3}\omega~,\\
\Omega_0^T &=&-4N_c N_f T\int\frac{d^3p}{(2\pi)^3}\log\left(1+e^{-\beta\omega}\right)~.
\end{eqnarray}
The valence quark contribution is finite and is not affected by renormalization. 
Therefore we focus on the renormalization of the vacuum part.
Within a $3-$momentum UV cutoff we find, in the limit $\Lambda \gg m_q$ and neglecting terms which
do not depend on the quark condensate,
\begin{equation}
\frac{\Omega_0^0}{N_c N_f} = -\frac{m_q^4}{32\pi^2} -\frac{m_q^2 \Lambda^2}{4\pi^2}
+\frac{m_q^4}{8\pi^2}\log\frac{2\Lambda}{m_q}~.
\label{eq:r1}
\end{equation} 
In order to remove the UV divergences we add the two counterterms to the thermodynamic potential,
\begin{equation}
\Omega^{c.t.} = \frac{\delta\lambda}{4}\frac{m_q^4}{g^4} + \frac{\delta v}{2}\frac{m_q^2}{g^2}~,
\label{eq:Vct}
\end{equation}
and impose the renormalization conditions~\cite{Suganuma:1990nn,Frasca:2011zn}
\begin{equation}
\left.\frac{\partial (\Omega_0^0 + \Omega^{c.t.})}{\partial m_q}\right|_{m_q = g f_\pi}=
\left.\frac{\partial^2 (\Omega_0^0 + \Omega^{c.t.})}{\partial m_q^2}\right|_{m_q = g f_\pi}=0~,
\label{eq:r2}
\end{equation}
which amount to the requirement that the one-loop contributions,
represented by $\Omega_0$, do not affect the expectation value
of the scalar field and the mass of the scalar meson. 
After elementary evaluation of the counterterms we find, 
putting $\Omega_0^{ren} \equiv \Omega_0^0 + \Omega^{c.t.}$,
\begin{equation}
\frac{\Omega_0^{ren}}{N_c N_f} = -\frac{g^2 f_\pi^2}{8\pi^2}m_q^2 
+ \frac{3}{32\pi^2}m_q^4
-\frac{m_q^4}{8\pi^2}\log\frac{m_q}{g f_\pi}~.
\label{eq:r3}
\end{equation}

In this article we are interested to build up the GL expansion
at the chiral critical line. To this end we need to expand the thermodynamic potential,
$\Omega_0^{ren} + \Omega_0^T$, in powers of $m_q$ around $m_q=0$.
An inspection of Eq.~\eqref{eq:r3} reveals that in the zero temperature part of the
thermodynamic potential a nonanalytic term, proportional to $m_q^4\log m_q$, is present.
However this is not a problem; in fact, as discussed in~\cite{Skokov:2010sf},
summing up the thermal part of the potential removes the dependence on the 
logarithm of the quark mass, leaving an analytic function which can be expanded
around $m_q = 0$. In order to show how this cancellation occurs 
we take advantage of the Dolan-Jackiw large temperature expansion~\cite{Dolan:1973qd}, 
see~\cite{Quiros:1999jp} for a review,
\begin{equation}
\frac{\Omega_0^T}{N_c N_f} =-\frac{7\pi^2 T^4}{180}+ \frac{m_q^2 T^2}{12} + 
\frac{m_q^4}{8\pi^2}\left[\log\frac{m_q}{\pi T} + \gamma_E - \frac{3}{4}\right] 
+ {\cal O}\left(m_q^6/T^6\right)~.
\label{eq:r7}
\end{equation}
Summing Eq.~\eqref{eq:r3}, which is
valid at any temperature, to Eq.~\eqref{eq:r7},
which is valid only in the limit $T \gg m_q$ (which is the limit which we are interested to)
results in the cancellation of the $\log m_q$ term; taking into account the meson potential
Eq.~\eqref{eq:L2}, we find
\begin{equation}
\Omega_0^{ren} + \Omega_0^T + U= \frac{\alpha_2^R}{2}m_q^2 + \frac{\alpha_4^R}{4}m_q^4
+ {\cal O}\left(m_q^6/T^6\right)~,
\label{eq:m1}
\end{equation}
with
\begin{eqnarray}
\alpha_2^R &=& 2 N_c N_f\left(\frac{T^2}{12} - \frac{g^2 f_\pi^2}{8\pi^2}\right) - \frac{\lambda v^2}{g^2}~,
\label{eq:p1}\\
\alpha_4^R &=& 4 N_c N_f\left(\frac{\gamma_E - 3/4}{8\pi^2} +\frac{3}{32\pi^2}
   - \frac{1}{8\pi^2}\log\frac{\pi T}{g f_\pi}\right) + \frac{\lambda}{g^4}~,
\label{eq:p2}
\end{eqnarray}
and the superscript $R$ stands for renormalized. In the above equation, $\gamma_E$ corresponds
to the Euler-Mascheroni constant.

According to the general GL theory of phase transitions, the critical temperature is obtained
as a solution of the equation $\alpha_2^R = 0$, that is
\begin{equation}
T_c^2 = \frac{6\lambda v^2}{g^2 N_c N_f}+ \frac{3g^2 f_\pi^2}{2\pi^2}~.
\label{eq:m2}
\end{equation}
For numerical estimations we take the parameters of~\cite{Mizher:2010zb}, namely 
$\lambda=20$, $v=f_\pi$ and $g=3.3$; with this parameter set we find $T_c \approx 173$ MeV.
Moreover $\alpha_4^R(T_c) > 0$, which implies that the phase transition is of the second order.

\subsection{Ultraviolet convergence of $\delta\Omega$}
Here we summarize the discussion of~\cite{Frasca:2011zn} which proves the ultraviolet convergence
of $\delta\Omega$ in Eq.~\eqref{eq:PP}. Since the valence quark contribution is obviously finite,
we limit ourselves to consider the $T=0$ part in $\delta\Omega$. 
In $\Omega_B$ the zero temperature contribution is
\begin{equation} 
-N_c\sum_f \frac{|Q_f eB|}{2\pi}\sum_{n=0}^\infty\beta_n \int_{-\infty}^{+\infty}\frac{dp_z}{2\pi} E~;
\end{equation}
following~\cite{Frasca:2011zn} we regulate the above term 
introducing the function, ${\cal V}(s)$, of a complex
variable, $s$, as
\begin{equation}
{\cal V}(s) = -N_c\sum_f\frac{|Q_f
eB|}{2\pi}\sum_{n=0}^\infty\beta_n
\int_{-\infty}^{+\infty}\frac{dp_z}{2\pi} \left(p_z^2 + 2|Q_f eB|n +
m_q^2\right)^\frac{1-s}{2}~.  \label{eq:Vfs}
\end{equation}
The function ${\cal V}(s)$ can be analytically continued to $s=0$.
We define then $\Omega_{B}(T=0) =
\lim_{s\rightarrow 0^+} {\cal V}(s)$. After elementary integration
over $p_z$, summation over $n$ and taking the limit $s\rightarrow
0^+$, we obtain the result
\begin{eqnarray}
\Omega_B(T=0) &=&
N_c\sum_f\frac{m_q^4}{16\pi^2}\left(\frac{2}{s}-\log(2|Q_f eB|)+a\right)\nonumber\\
&&+N_c\sum_f\frac{|Q_f eB| m_q^2}{8\pi^2}\log\frac{m_q^2}{2|Q_f
eB|} \nonumber
\\
&&-N_c\sum_f \frac{(Q_f
eB)^2}{2\pi^2}\zeta^\prime\left(-1,q\right)~. \label{eq:F4bis}
\end{eqnarray}
where we have subtracted terms which do not depend explicitly on
the condensate. In the above equation, $\zeta\left(t,q\right)$ is
the Hurwitz zeta function; 
we have defined $q = (m_q^2
+ 2|Q_f eB|)/2|Q_f eB|$; furthermore, $a = 1 - \gamma_E -
\psi(-1/2)$, where $\gamma_E$ is the Euler-Mascheroni number and
$\psi$ is the digamma function. The derivative $\zeta^\prime
\left(-1,q\right) =d\zeta(t,q)/d t$ is understood to be computed
at $t=-1$. 

It is shown in~\cite{Frasca:2011zn} that 
the ultraviolet divergence of the lowest Landau level (LLL) contribution is
canceled by an analogous divergence coming from the higher Landau levels. 
Nevertheless a divergence still remains, arising from the higher Landau levels,
and that is represented by the explicit
$1/s$ pole in Eq.~\eqref{eq:F4bis}, which survives in the $\bm B\rightarrow 0$ limit 
and is obviously related to the divergence of the vacuum contribution
which we have examined in the previous Section.
The structure of the divergence in Eq.~\eqref{eq:F4bis}
is identical to that obtained within the dimensional regularization scheme; 
the apparent missing scale of the logarithm is hidden in the $1/s$ term, and appears
explicitly when the divergence is subtracted. Such a divergence
affects only the zero field, zero temperature thermodynamic potential; 
the corrections due to the magnetic field are either finite or independent on the
condensate~\footnote{In~\cite{Endrodi:2013cs} a term proportional
to $B^2/s$ is considered; this is a pure field contribution, which needs to
be renormalized and contributes to the total pressure. 
On the other hand, it is not coupled to the quark condensate; 
for this reason we do not take into account such a term
in the present study.}. 
As a matter of fact, taking the zero magnetic field limit
we find
\begin{equation}
\Omega_0(T=0)=
N_c N_f \frac{m_q^4}{16\pi^2}\left(\frac{2}{s}-\log m_q^2+a +
\frac{1}{2}\right)~; \label{eq:las}
\end{equation}
comparing Eqs.~\eqref{eq:F4bis} and~\eqref{eq:las} it is easily proved that the pole
$2/s$ is cancelled in the difference $\Omega_B(T=0) - \Omega_0(T=0)$,
\begin{eqnarray}
\delta\Omega(T=0) &=& -N_c\sum_f\left(\frac{m_q^4}{16\pi^2} + \frac{|Q_f eB|
m_q^2}{8\pi^2}\right)\log\frac{2|Q_f eB|}{m_q^2} \nonumber \\
&&-N_c\sum_f\frac{|Q_f eB|^2}{2\pi^2}\zeta^\prime\left(-1,q\right)
-N_c N_f\frac{m_q^4}{32\pi^2}~, \label{eq:uq}
\end{eqnarray}
hence proving the statement that $\delta\Omega$ is ultraviolet finite.
Since $\delta\Omega$ is finite, it does not depend on the actual regularization scheme.
In fact, in the following Section we are going to use a different regularization scheme
to compute the Ginzburg-Landau coefficients in the effective action for $m_q$ at the
critical line, employing a $3-$momentum cutoff, which is certainly less elegant
than the regularization scheme used in this Section,
but easily manageable analytically. In that case we will prove excplicitly
the cancellation of ultraviolet divergences in the relevant coefficients 
of the GL expansion.

Before going ahead, it is useful to remind that our goal in this article is to
obtain analytically the expression of the thermodynamic potential at the critical line,
within an expansion in powers of $m_q$. However, we notice that the argument of the log in Eq.~\eqref{eq:uq} forbids
the analytic expansion of the thermodynamic potential in powers of $m_q/|eB|$. 
This problem is analogous to the one encountered in the $\bm B=0$ case, see for example
Ref.~\cite{Skokov:2010sf} for a nice discussion; it is well known that in the latter case,
summing the valence quark contribution to the thermodynamic potential results in 
the replacement of the $m_q/\Lambda$ argument in the log in the zero temperature term 
with a $T/\Lambda$ argument, turning the thermodynamic potential to an analytic function
of $m_q$ at $m_q=0$, then permitting the $m_q-$power expansion.   
In the case of finite $\bm B$ we face the same problem; as we will show explicitly in the
next Section, this is cured by summing the finite temperature contribution to the
thermodynamic potential.

\section{Effective action at the critical line}
In this Section we present the novelty of our study. Our goal is to
expand $\Omega$ in powers of $m_q$, in order to build up the
effective potential at the critical line for the order parameter. 
In particular, we compute the quadratic and the quartic coefficients
of the Ginzburg-Landau expansion in Eq.~\eqref{eq:i1}, putting 
\begin{equation}
\alpha_i \equiv \alpha_i^R + \delta\alpha_i~.
\label{eq:da}
\end{equation}
Here $\alpha_i^R$ and $\delta\alpha_i$ are computed from $\Omega_0$ and $\delta\Omega$
in Eq.~\eqref{eq:PP}, respectively, as $\alpha_i^R = \partial^i\Omega_0/\partial m_q^i|_{mq=0}$
and $\delta\alpha_i = \partial^i(\Omega_B - \Omega_0)/\partial m_q^i|_{mq=0}$. 
The zero field coefficients have been computed in the previous Section, 
see Eqs.~\eqref{eq:p1} and~\eqref{eq:p2}, and are the ones which are affected by
the renormalization procedure. Since $\Omega_B - \Omega_0$ is finite,
it is not sensitive to renormalization. 
In this Section we compute $\delta\alpha_2$ and $\delta\alpha_4$. 

As we have stressed in the previous Section, the quantity $\Omega_B - \Omega_0$ is finite 
both in the ultraviolet (UV) and in the infrared (IR). However, in the intermediate steps
of the computations, when the coefficients from $\Omega_0$ and $\Omega_B$ are computed
independently, it is necessary to regulate the several addenda both in the UV and in
the IR. We achieve this by introducing the $3-$momentum cutoffs $\Lambda$ and 
$\varepsilon$ respectively. However, we will prove explicitely that the final result does not depend on these cutoffs, since
both the UV and the IR divergences are removed at any order in $m_q$ when the vacuum contribution
is subtracted from the finite field one, as we have discussed in the previous Section. 

In the computation of $\delta\alpha_i$ we find convenient use the following decomposition 
of $\Omega_B$:
\begin{equation}
\Omega_B = \Omega_{B,0}^{LLL} + \Omega_{B,T}^{LLL} +
\Omega_{B,0}^{hLL} + \Omega_{B,T}^{hLL}~,
\label{eq:deco} 
\end{equation}
where the superscripts $LLL$, $hLL$ correspond to lowest Landau level and higher Landau levels,
respectively; moreover, the subscripts $0$ and $T$ correspond to the zero temperature
and valence quark contributions respectively. By means of the above decomposition it is easy to 
identify the several contributions to each GL coefficient, as well as to check
how the intermediate steps divergences combine to produce a finite result.

\subsection{Computation of $\delta\alpha_2$}
According to the definitions in Eq.~\eqref{eq:da}, in order to compute $\delta\alpha_i$
we need to expand both $\Omega_B$ and $\Omega_0$. 
The computation of the zero field contribution is straightforward; we find
\begin{equation}
\Omega_0 = \frac{m_q^2}{2}\left[
\frac{N_c N_f}{\pi^2}
\left(
\frac{\pi^2 T^2}{6} - \frac{\Lambda^2}{2}
\right)\right]
~,~~~\text{at the order $m_q^2$}~,
\label{eq:V1}
\end{equation}   
where $\Lambda$ is an intermediate-step UV cutoff which is introduced to regularize the result;
as we have already stressed, when we subtract the above equation to the finite field contribution, 
the divergence is exactly cancelled, leaving a finite result.

Taking the second derivative of $\Omega_{B,0}^{LLL} + \Omega_{B,T}^{LLL}$ with respect to $m_q$ at $m_q=0$ we find
\begin{equation}
\Omega_{B,0}^{LLL} + \Omega_{B,T}^{LLL} = \frac{m_q^2}{2}
\left[-N_c\sum_f\frac{|Q_f e B|}{2\pi^2}\left(
\log\frac{\Lambda}{T} + a_2
\right)\right]
~,~~~\text{at the order $m_q^2$}~,
\label{eq:V2}
\end{equation}
with $a_2 \approx 0.139$ resulting from a convergent numerical integral. The above result
is obtained by means of elementary integrations. The only care needed in the computation
is to regulate the contributions coming from $\Omega_{B,0}^{LLL}$ and $\Omega_{B,T}^{LLL}$
in the IR, introducing an IR cutoff, $\varepsilon$; summing the finite temperature contribution
to the zero temperature one, $\lambda$ cancels exactly leaving a finite IR result.

The computation involving the higher Landau levels is complicated by the summation over 
the infinite tower of levels. In order to regulate the UV divergences in the intermediate steps
of the computation we use the scheme of~\cite{Fukushima:2010fe}, requiring that
$p_z^2 + 2|Q_f eB|n \leq \Lambda^2$ with $\Lambda$ an UV cutoff.
Then the zero temperature contribution reads 
\begin{equation}
\Omega_{B,0}^{hLL} = \frac{m_q^2}{2}\left[
-N_c\sum_f \frac{|Q_f eB|}{\pi^2}\sum_{n=1}^{N_\Lambda}
\log\frac{\sqrt{N_\Lambda -n} + \sqrt{N_\Lambda}}{\sqrt{n}}\right]
~,~~~\text{at the order $m_q^2$}~,
\label{eq:V4_in}
\end{equation}
where $N_\Lambda = \Lambda^2/(2 |Q_f eB|)$. In the above equation the summation over Landau levels
has to be performed.
The divergent part can be extracted analytically by using the
Euler-McLaurin formula, 
\begin{equation}
\sum_{n=1}^{N_{\Lambda}} f(n) \approx \int_{1}^{N_\Lambda}f(z)dz 
+ \frac{f(1) + f(N_\Lambda)}{2}~,
\label{eq:EM}
\end{equation}
while the finite part has to be computed numerically.
We find, in the limit $\Lambda^2 \gg eB$,
\begin{equation}
\sum_{n=1}^{N_\Lambda}
\log\frac{\sqrt{N_\Lambda -n} + \sqrt{N_\Lambda}}{\sqrt{n}} \approx 
\frac{\Lambda^2}{2|Q_f eB|} + \frac{1}{4}\log\frac{2|Q_f eB|}{\Lambda^2} - b_2~,
\end{equation}
with $b_2\approx 0.806$ independent on the fermion charge, hence leading to
\begin{equation}
\Omega_{B,0}^{hLL} = \frac{m_q^2}{2}\left[-N_c\sum_f\frac{|Q_f eB|}{\pi^2}
\left(
\frac{\Lambda^2}{2|Q_f eB|} + \frac{1}{4}\log\frac{2|Q_f eB|}{\Lambda^2} - b_2
\right)
\right]
~,~~~\text{at the order $m_q^2$}~.
\label{eq:V4}
\end{equation}
Comparing the above equation with Eqs.~\eqref{eq:V1} and~\eqref{eq:V2} we find that
the UV divergences are perfectly cancelled, as it should.

The last computation is the thermal contribution of the higher Landau levels. This computation
is a bit lengthy but straightforward. Defining
\begin{equation}
Y_n = -\frac{2T}{\pi^2}
\sum_f |Q_f eB|
\int_0^\infty dp_z\frac{\partial^2}{\partial m_q^2}
\left.
\log\left(1+e^{-\beta E}\right)
\right|_{m_q = 0}~,~~~n\geq 1~,
\end{equation}
we have
\begin{equation}
\Omega_{B,T}^{hLL} = \frac{m_q^2}{2} N_c
\sum_{n=1}^{\infty} Y_n 
~,~~~\text{at the order $m_q^2$}~;
\label{eq:V5_in}
\end{equation}
in order to easily combine Eq.~\eqref{eq:V5_in} with Eq.~\eqref{eq:V1} and~\eqref{eq:V2} 
we define $Y=\sum_n Y_n$ and
\begin{equation}
F_2=Y - \frac{T^2}{6} + \frac{|Q_f eB|}{4\pi^2}\log\frac{T^2}{2|Q_f e B|}~,
\label{eq:F2_def}
\end{equation}
in such a way
\begin{equation}
\Omega_{B,T}^{hLL} = \frac{m_q^2}{2}N_c\sum_f \left(
\frac{T^2}{6} + \frac{|Q_f eB|}{4\pi^2}\log\frac{2|Q_f e B|}{T^2}
 +F_2
\right) 
~,~~~\text{at the order $m_q^2$}~.
\label{eq:V5}
\end{equation}
We have not been able to obtain an analytic expression for $F_2$; on the other hand,
just on the base of dimensional analysis and on the observation that $F_2\rightarrow 0$
in the $\bm B\rightarrow 0$ limit, we expect $F_2 = (|Q_f eB|/\pi^2)\times {\cal G}_2(|Q_f eB|/T^2)$
(the overall $1/\pi^2$ is introduced just for convenience).
In the case $|eB|/T^2 \ll 1$ the function ${\cal G}_2$ can then be determined by a best fit procedure;
in the opposite limit $|eB|/T^2 \gg 1$ the thermal contribution of the higher Landau levels
is suppressed, which implies $Y \rightarrow 0$ in Eq.~\eqref{eq:F2_def}. Then we find respectively
($x\equiv |eB|/T^2$)
\begin{eqnarray}
{\cal G}_2 &\approx& -13.51 -7.24\log x~,~~~x \ll 1\\
{\cal G}_2 &\approx& -\frac{1}{4}\log2x
-\frac{\pi^2}{6x}~,~~~x \gg 1~.
\end{eqnarray}

Summing up all the finite $\bm B$ contributions and subtracting the $\bm B=0$ one we thus find
\begin{equation}
\delta\alpha_2 =  N_c \sum_f\frac{|Q_f eB|}{\pi^2}\left[
-\frac{1}{2}a_2
+ {\cal G}_2\left(\frac{|Q_f eB|}{T^2}\right) +b_2
\right]~.
\label{eq:FV1}
\end{equation}

\subsection{Computation of $\delta\alpha_4$}
From $\Omega_0$ we find
\begin{equation}
\Omega_0 = \frac{m_q^4}{4}\left(\frac{3N_c N_f}{2\pi^2}\log\frac{\Lambda^2}{T^2} - N_c N_f c_4^0\right)
~,~~~\text{at the order $m_q^4$}~,
\label{eq:J1}
\end{equation}   
where $c_4^0\approx 0.265$ resulting from a finite numerical integral, and $\Lambda$ is
an ultraviolet cutoff that we introduce to cut the $3-$momentum integral. At the end of the
calculation, the log divergence will be cancelled by the finite $\bm B$ contribution, 
in agreement with the argument given in the previous Section.

Taking the fourth derivative of $\Omega_{B,0}^{LLL} + \Omega_{B,T}^{LLL}$ 
with respect to $m_q$ and computing it at $m_q=0$ we find
\begin{equation}
\Omega_{B,0}^{LLL} + \Omega_{B,T}^{LLL} = \frac{m_q^4}{4}\left[
N_c\sum_f\frac{|Q_f e B|}{\pi^2}\frac{3}{T^2}
a_4
\right]
~,~~~\text{at the order $m_q^4$}~,
\label{eq:J2}
\end{equation}
In order to obtain Eq.~\eqref{eq:J2} we have
used an ultraviolet $3-$momentum cutoff regularization, then sent the UV cutoff to infinity being the result
finite in the ultraviolet. 
In the above equation $a_4 \approx 0.11$ results from a convergent numerical integral.

As for $\delta\alpha_2$, the computation involving the higher Landau levels is complicated by the summation over 
the infinite levels. In order to regulate the UV divergences in the intermediate steps
of the computation we use again the regulation scheme which has lead to Eq.~\eqref{eq:V4_in}.
Then the zero temperature contribution reads 
\begin{equation}
\Omega_{B,0}^{hLL} = \frac{m_q^4}{4}
N_c\sum_f \frac{3}{2\pi^2}\sum_{n=1}^{N_\Lambda}
\frac{\sqrt{\Lambda^2 -2|Q_f eB|n}}{n\Lambda}
~,~~~\text{at the order $m_q^4$}~,
\label{eq:J4_in}
\end{equation}
where $N_\Lambda = \Lambda^2/(2 |Q_f eB|)$. In the above equation the summation over Landau levels
has to be performed; the divergent contribution can be extracted once again
by virtue of Eq.~\eqref{eq:EM}; the finite part is then computed numerically.
We find,  in the limit $\Lambda^2 \gg eB$, 
\begin{equation}
\sum_{n=1}^{N_\Lambda}
\frac{\sqrt{\Lambda^2 -2|Q_f eB|n}}{n\Lambda} \approx 
\log\frac{\Lambda^2}{2|Q_f eB|} - d_4~,
\end{equation}
with $d_4\approx 0.0360$, hence leading to
\begin{equation}
\Omega_{B,0}^{hLL} = \frac{m_q^4}{4}
N_c\sum_f \frac{3}{2\pi^2}\left(
\log\frac{\Lambda^2}{2|Q_f eB|} - d_4\right)
~,~~~\text{at the order $m_q^4$}~.
\label{eq:J4}
\end{equation}
The above equation is quite useful since 
it allows to verify easily that once Eq.~\eqref{eq:J1} is subtracted, the UV divergence 
is cancelled leaving a finite final result.

The last computation is the thermal contribution of the higher Landau levels. As in the case
of $\delta\alpha_2$ the computation is a bit lenghty but straightforward. Firstly we define
\begin{equation}
J_n = -\frac{2T}{\pi^2}
\sum_f |Q_f eB|
\int_0^\infty dp_z\frac{\partial^4}{\partial m_q^4}
\left.
\log\left(1+e^{-\beta E}\right)
\right|_{m_q = 0}~,~~~n\geq 1~;
\end{equation}
then we have
\begin{equation}
\Omega_{B,T}^{hLL} = \frac{m_q^4}{4} N_c
\sum_{n=1}^{\infty} J_n 
~,~~~\text{at the order $m_q^4$}~;
\label{eq:J5_in}
\end{equation}
in order to easily combine subtract Eq.~\eqref{eq:J5_in} with Eq.~\eqref{eq:J1} and~\eqref{eq:J4} 
we define $J=\sum_n J_n$ and
\begin{equation}
F_4=J + \frac{3}{2\pi^2}\log\frac{T^2}{2|Q_f e B|}~,
\label{eq:F4_def}
\end{equation}
in such a way
\begin{equation}
\Omega_{B,T}^{hLL} = \frac{m_q^4}{4}N_c\left(
- \frac{3}{2\pi^2}\sum_f\log\frac{T^2}{2|Q_f e B|} +\sum_f F_4
\right) 
~,~~~\text{at the order $m_q^4$}~.
\label{eq:J5}
\end{equation}
As for the case of $F_2$, we have not been able to obtain an analytic expression for $F_4$
for any value of the magnetic field strength; we have limited ourselves to verify numerically
that in the $\bm B\rightarrow 0$ limit we have $F_4\rightarrow -c_4^0 + 3 d_4/2\pi^2$ independently on the fermion charge, 
which guarantees $\delta\alpha_4\rightarrow 0$ in the same limit.
In the case $|eB|/T^2 \ll 1$ a best fit procedure can be used to extract the $\bm B-$dependence
of $F_4$ on the magnetic field strength; on the other hand, in the case $|eB|/T^2 \gg 1$
the finite temperature contribution to the thermodynamic potential is dominated by the lowest Landau level,
which implies $J\rightarrow 0$ in Eq.~\eqref{eq:F4_def}. We find
\begin{eqnarray}
F_4 &\approx& \frac{3}{2\pi^2}d_4 - c_4^0~,~~~x \ll 1\\
F_4 &\approx& -\frac{3}{2\pi^2}\log2x~,~~~x \gg 1~.
\end{eqnarray}

Summing up all the finite $\bm B$ contributions and subtracting the $\bm B=0$ one we find
\begin{equation}
\delta\alpha_4  =N_c \sum_f\left(
\frac{3a_4}{\pi^2}\frac{|Q_f e B|}{T^2}
-\frac{3}{2\pi^2}d_4
+ F_4 + c_4^0
\right)~.
\label{eq:F1}
\end{equation}

\begin{figure}[t!]
\begin{center}
\includegraphics[width=7cm]{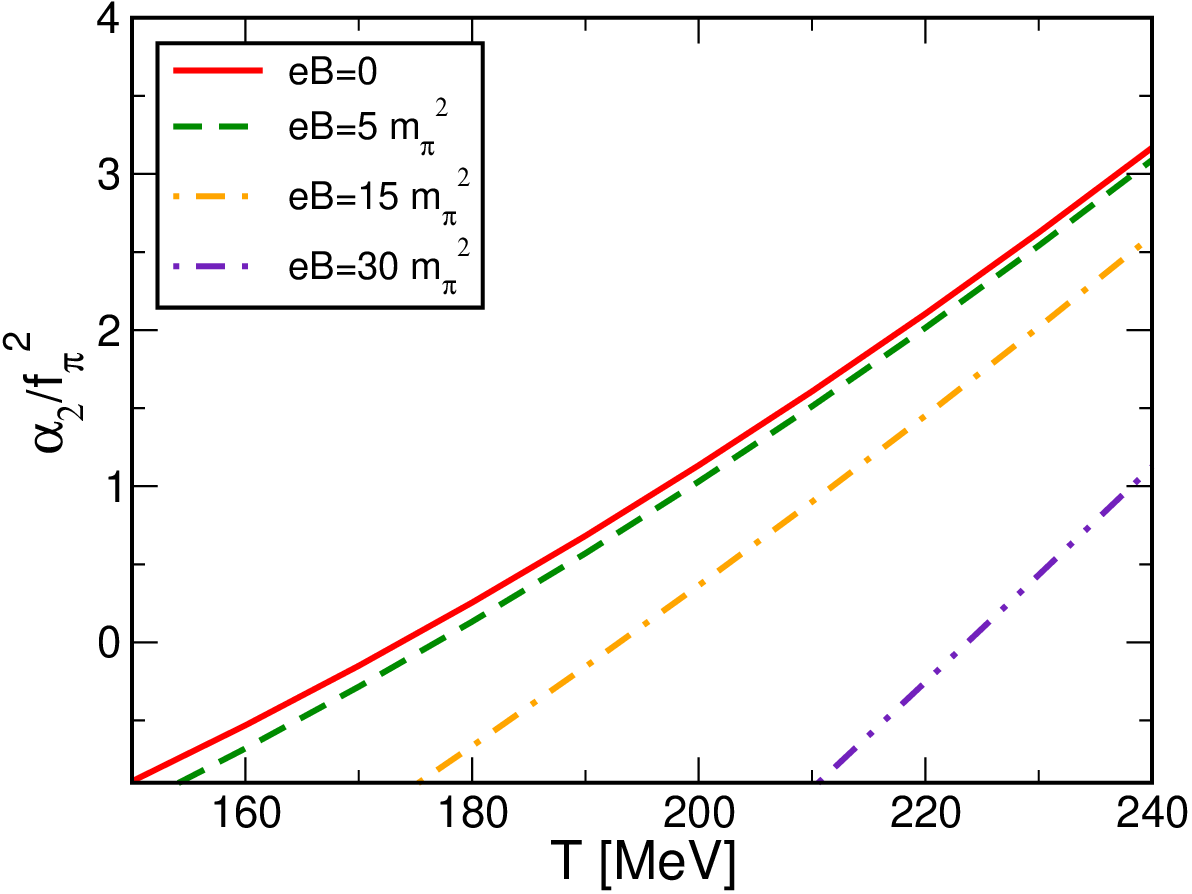}~~~
\includegraphics[width=7cm]{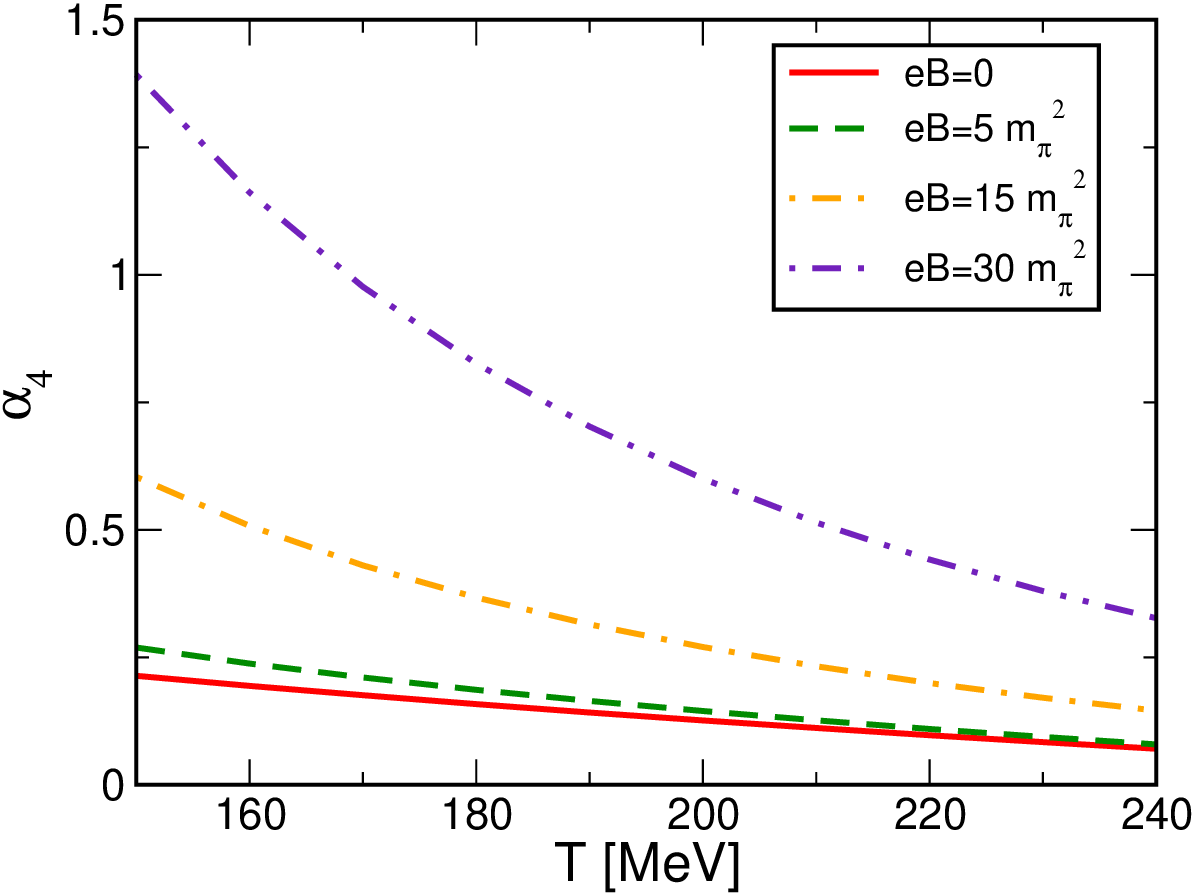}
\caption{\label{Fig:a1}\small Coefficients $\alpha_2$ ({\em left panel}) 
and $\alpha_4$ ({\em right panel}) as a function of temperature, for
several values of the magnetic field strength.}
\end{center}
\end{figure}

In Fig.~\ref{Fig:a1} we plot the coefficients $\alpha_2$  
and $\alpha_4$ as a function of temperature, for
several values of the magnetic field strength. The location of the zeros of the quadratic
coefficient is affected by the magnetic field: the larger $eB$, the larger the
critical temperature. This model prediction is in disagreement with the
recent lattice simulations of~\cite{Bali:2011qj}, but understanding the reason
of the discrepancy is beyond the scope of our study.
At a fixed temperature, the magnetic field increases the numerical value 
of $\delta\alpha_4$ as well, which might sound unexpected since increasing $\bm B$
should finally result in making the phase transition closer to a first order, hence lowering the value
of $\alpha_4$.

\begin{figure}[t!]
\begin{center}
\includegraphics[width=7cm]{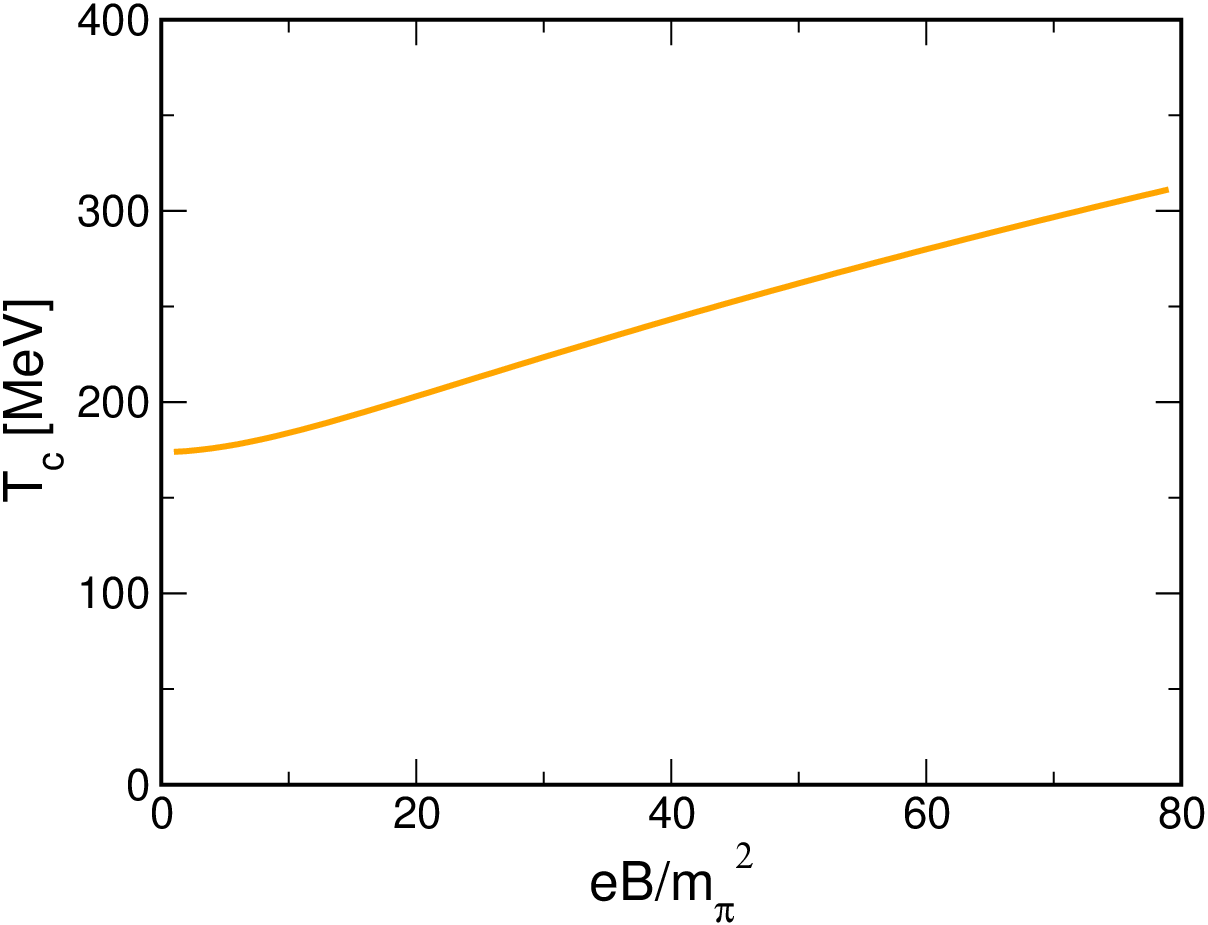}~~~
\includegraphics[width=7cm]{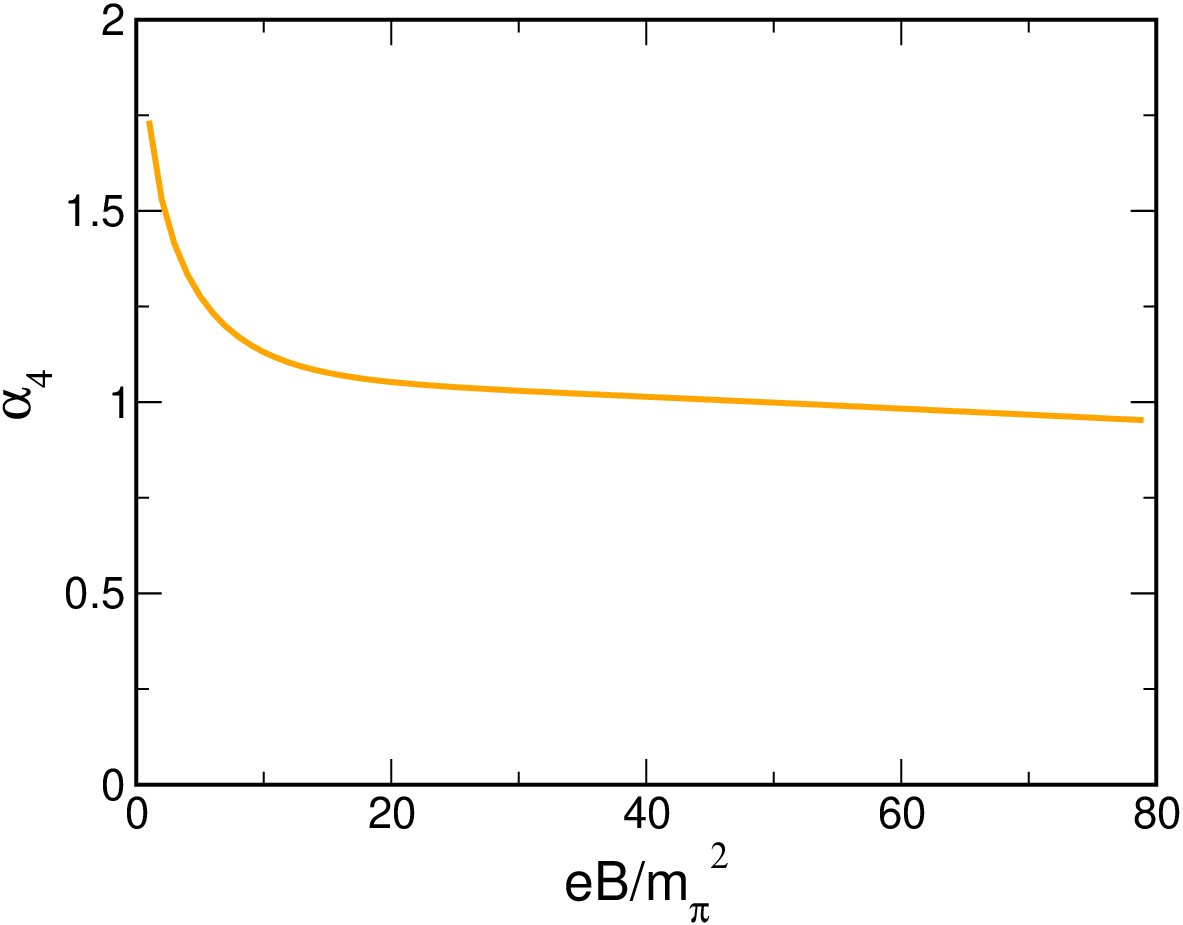}
\caption{\label{Fig:a3}\small Critical temperature ({\em left panel}) 
and $\alpha_4$ at the critical temperature ({\em right panel}) as a function of $eB/m_\pi^2$, 
for the case of the renormalized model. }
\end{center}
\end{figure}

In order to understand why the magnetic field enhances the strength of the phase transition, 
we plot in Fig.~\ref{Fig:a3} the critical temperature (left panel)
and $\alpha_4$ at the critical temperature (right panel) as a function of $eB/m_\pi^2$, 
for the case of the renormalized model.  We find that when computed at $T_c$, the value of 
$\alpha_4$ is a decreasing function of $eB$, thus making the phase transition closer
to a first order. However, the $\alpha_4$
is never zero. This means that, at least within the magnetic field range we have analyzed in this study
(which is however well beyond the largest magnetic fields expected to be produced in heavy ion
collisions, as well as the one produced in the early universe at the QCD phase transition),
no critical point is present in the phase diagram, in agreement with the lattice simulations~\cite{Bali:2011qj}.

\subsection{Lowest Landau level approximation}
It is instructive to compute the renormalized coefficients of the GL effective action in the LLL approximation,
which is valid when $eB \gg  T^2$:
within the renormalized model, this amounts to consider only the LLL contribution in the thermal part of the $\bm B-$dependent
thermodynamic potential, that is $Y=0$ in Eq.~\eqref{eq:F2_def} and $J=0$ in Eq.~\eqref{eq:F4_def}.
On the other hand, one has to be careful in defining the
LLL limit in the vacuum term, because the hLLs are important to cancel the divergences arising from the expansion 
of $\Omega_0$ and of $\Omega_0^{LLL}$.

In the case of the second order coefficient,
the hLLs vacuum term combines non trivially with the
other terms, see Eq.~\eqref{eq:V4}; in particular, the quadratic divergence
cancels with the analogous divergence in $\Omega_0$, while the log-type divergence
combines with the same kind of divergence of the LLL to give a renormalized
result $\propto|Q_f e B|\log2|Q_f e B|/T^2$. Parametrically this is the leading contribution
to the second order coefficient in the limit $|eB| \gg T^2$,
and corresponds to the log-type term in the asymptotic expansion of ${\cal G}_2$
in Eq.~\eqref{eq:FV1}.

For the case of the quartic order coefficient, 
the log-type divergence in Eq.~\eqref{eq:J4} combines with the same kind of divergence
in Eq.~\eqref{eq:J1} to give a finite result $\propto\log2|Q_f e B|/T^2$.
However this contribution is parametrically subleading in comparison with the
finite LLL contribution in Eq.~\eqref{eq:J2} which is finite once we combine the
$T=0$ and the finite temperature terms, and it grows up as $|Q_f e B|/T^2$.
Moreover, parametrically $\alpha_2^R \ll \delta\alpha_2$ and $\alpha_4^R \ll \delta\alpha_4$
(we have verified numerically that this relation holds even for
not so large values of $|eB|/T^2$).
This observations based on the asymptotic behavior of the GL coefficients
are quite interesting, since they reveal that the critical temperature is
determined by a renormalized combination of the contribution of the LLL and the hLLs;
hence, both the LLL and the hLLs are effective to shift $T_c$.
However, the sign of the quartic coefficient is asymptotically determined
by the LLL only, meaning that only the LLL affects the strength of the phase transition
in the large field limit. 
Putting all together we thus find in the $|eB|/T^2 \gg 1$ limit (the symbol $\asymp$ corresponds to
asymptotic limit)
\begin{eqnarray}
\alpha_2 &\asymp& -N_c \sum_f\frac{|Q_f eB|}{4\pi^2}
\log\frac{2|Q_f eB|}{T^2}~,\label{eq:LLL2}
\\
\alpha_4 &\asymp& N_c \sum_f 
\frac{3a_4}{\pi^2}\frac{|Q_f e B|}{T^2}~.
\label{eq:LLL4}
\end{eqnarray} 
From Eq.~\eqref{eq:LLL4} we read that the quartic coefficient of the GL effective action
is positive in the large field limit, for every value of the temperature. This implies that 
the phase transition is of the second order at zero chemical potential, and this conclusion
is not affected by the strength of the magnetic field.

\section{Comparison with the nonrenormalized model}
It is interesting to compare the results summarized in Eqs.~\eqref{eq:FV1} and~\eqref{eq:F1}
with those obtained within a nonrenormalized model. If we perform the computation of
the thermodynamic potential in the latter case, 
the UV cutoff is kept finite and is considered a parameter, fixed to reproduce
some phenomenological quantity; moreover, renormalization of the thermodynamic potential
is not performed. Hence in this case the coefficients of the GL expansion 
at the critical line are given by the coefficients of the expansion of $\Omega_B$,
which will contain an explicit dependence on the UV cutoff (in the case of the
renormalized model, we add and subtract the zero field potential $\Omega_0$, which contains
the UV divergences, and renormalize the latter; the difference $\Omega_B - \Omega_0$ is UV finite
and does not need renormalization). 

The calculation of the GL coefficients in the nonrenormalized model
follow the same lines of those we have performed in the renormalized case.
We introduce a $3-$momentum UV cutoff, $\Lambda$, to regulate the divergence
of the zero temperature potential. The final results will depend on the numerical value
of $\Lambda$. 

The computation of $\alpha_2$ using the fixed cutoff scheme leads to the same 
results summarized in Eqs.~\eqref{eq:V2},~\eqref{eq:V4_in} and~\eqref{eq:V5_in}.
In the present case however we cannot use the asymptotic expression in Eq.~\eqref{eq:V4}
since cutoff $\Lambda$ is kept fixed and the condition
$\Lambda \gg |eB|$ might not apply. To the quark bubble, the contribution
from the meson potential equal to $- \lambda v^2/g^2$
has to be added, see Eq.~\eqref{eq:p1}.

For what concerns the computation of $\alpha_4$, the only formal difference 
is in the LLL contribution, which in this case reads
\begin{equation}
\Omega_{B,0}^{LLL}+\Omega_{B,T}^{LLL} = \frac{m_q^4}{4}\left[
N_c\sum_f\frac{|Q_f e B|}{\pi^2}\frac{3}{T^2}
\left(
a_4 - \frac{T^2}{4\Lambda^2}
\right)
\right]
~,~~~\text{at the order $m_q^4$}~,
\label{eq:H3}
\end{equation}
which agrees with Eqs.~\eqref{eq:LLL4} in the $\Lambda\rightarrow +\infty$ limit as expected.
The quartic coefficient becomes negative if the critical temperature, $T_c$, is 
larger than $T_c^\star = 2\Lambda\sqrt{a_4}\approx  0.66 \Lambda$.
Thus the change of the order of the phase transition seems inevitable within the
nonrenormalized model, at least in the case of very large magnetic field strenghts.
However this conclusion is very sensitive to the numerical value of the UV cutoff;
in the limit $\Lambda\rightarrow +\infty$, the contribution in Eq.~\eqref{eq:H3}
is positive definite, as it happens for the case of the renormalized model, see
Eq.~\eqref{eq:LLL4}.
The higher Landau levels contribution is still formally given by Eqs.~\eqref{eq:J4_in}
and~\eqref{eq:J5_in}; once again, we cannot use the asymptotic form in Eq.~\eqref{eq:J4}
because the UV cutoff is a fixed finite number. Finally, the contribution
from the meson potential equal to $\lambda/g^4$
has to be added, see Eq.~\eqref{eq:p2}.

\begin{figure}[t!]
\begin{center}
\includegraphics[width=7cm]{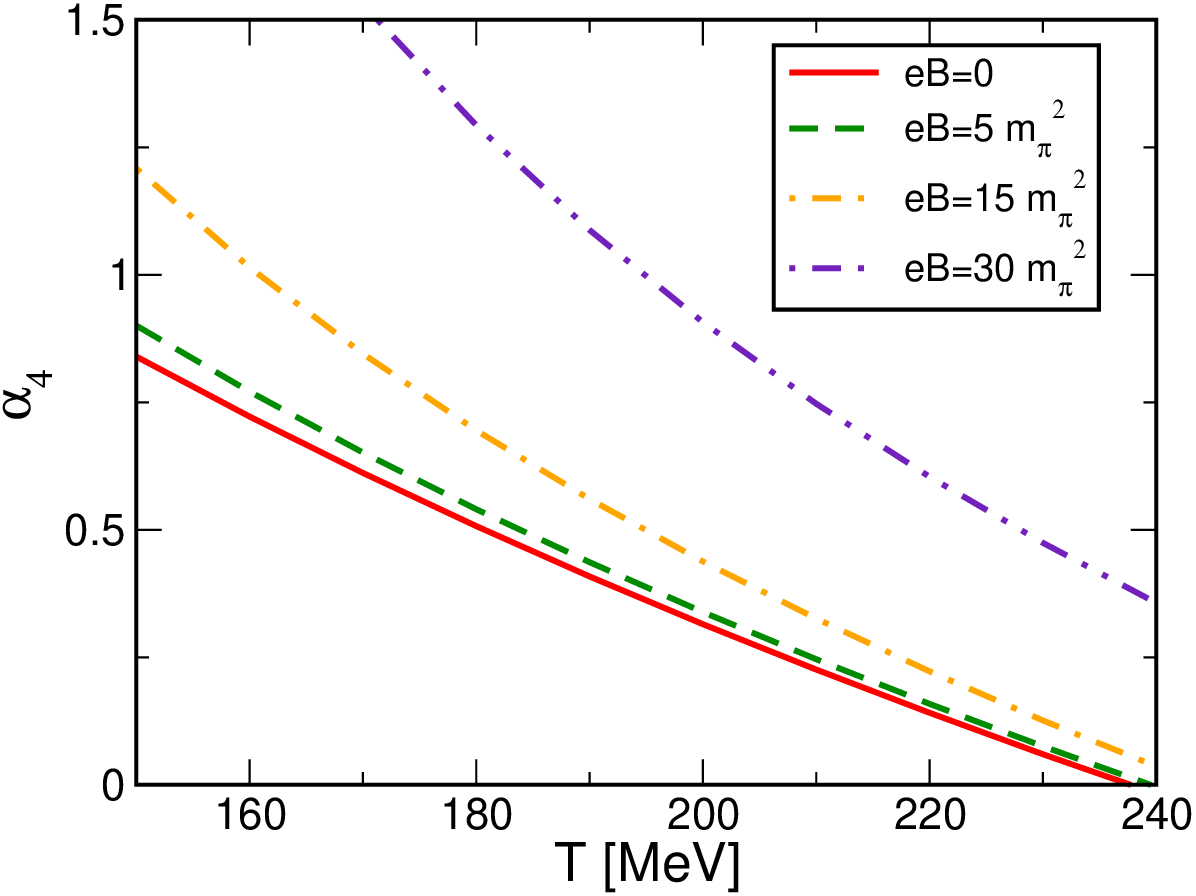}
\caption{\label{Fig:a2}\small Coefficient $\alpha_4$ as a function of temperature, for
several values of the magnetic field strength, in the case of the nonrenormalized model.}
\end{center}
\end{figure}

In Fig.~\ref{Fig:a2} we plot the coefficients  $\alpha_4$ as a function of temperature, for
several values of the magnetic field strength, in the case of the nonrenormalized model.
We do not plot $\alpha_2$ since its behavior does not change qualitatively switching from
the renormalized model to the nonrenormalized one. 
The numerical values of the parameters are taken by Ref.~\cite{Frasca:2011zn},
and are $\lambda= 4.67$, $\Lambda=560$ MeV and $v^2 = -1.8 m_q^2$ with $m_q = g f_\pi = 335$ MeV.
In this case the $v^2$ is negative, since the nonrenormalized quark bubble 
is taken into account when the requirement $\partial(\Omega_B + U)/\partial\sigma = 0$
for $\sigma = f_\pi$ at $\bm B=0$ and $T=0$.
As in the case of the renormalized model analyzed in the previous Section, 
increasing the value of the magnetic field strength at fixed temperature leads
to an increasing of $\alpha_4$. This behavior seems to be counterintuitive, since the 
hardening of the phase transition with the magnetic field, and eventually the
change of the transition from second order to first order, 
should occur because of a change of sign of $\alpha_4$; hence, one would expect
that the larger the magnetic field strength, the smaller $\alpha_4$.  However,
this naive argument ignores the possibility that the critical temperature
increases with the magnetic field strength, and that the GL expansion
is quantitatively reliable only in the very proximity of the phase transition.

\begin{figure}[t!]
\begin{center}
\includegraphics[width=7cm]{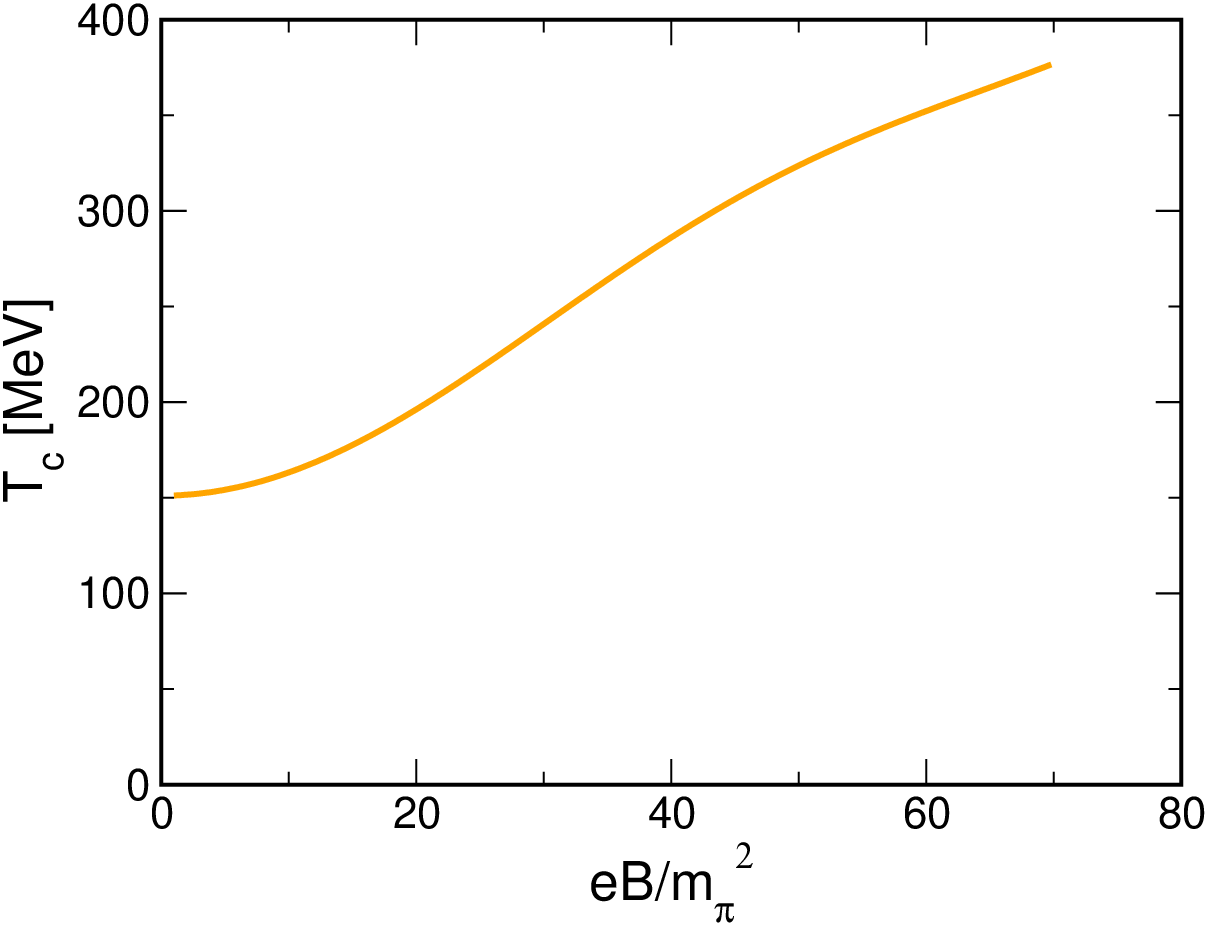}~~~
\includegraphics[width=7cm]{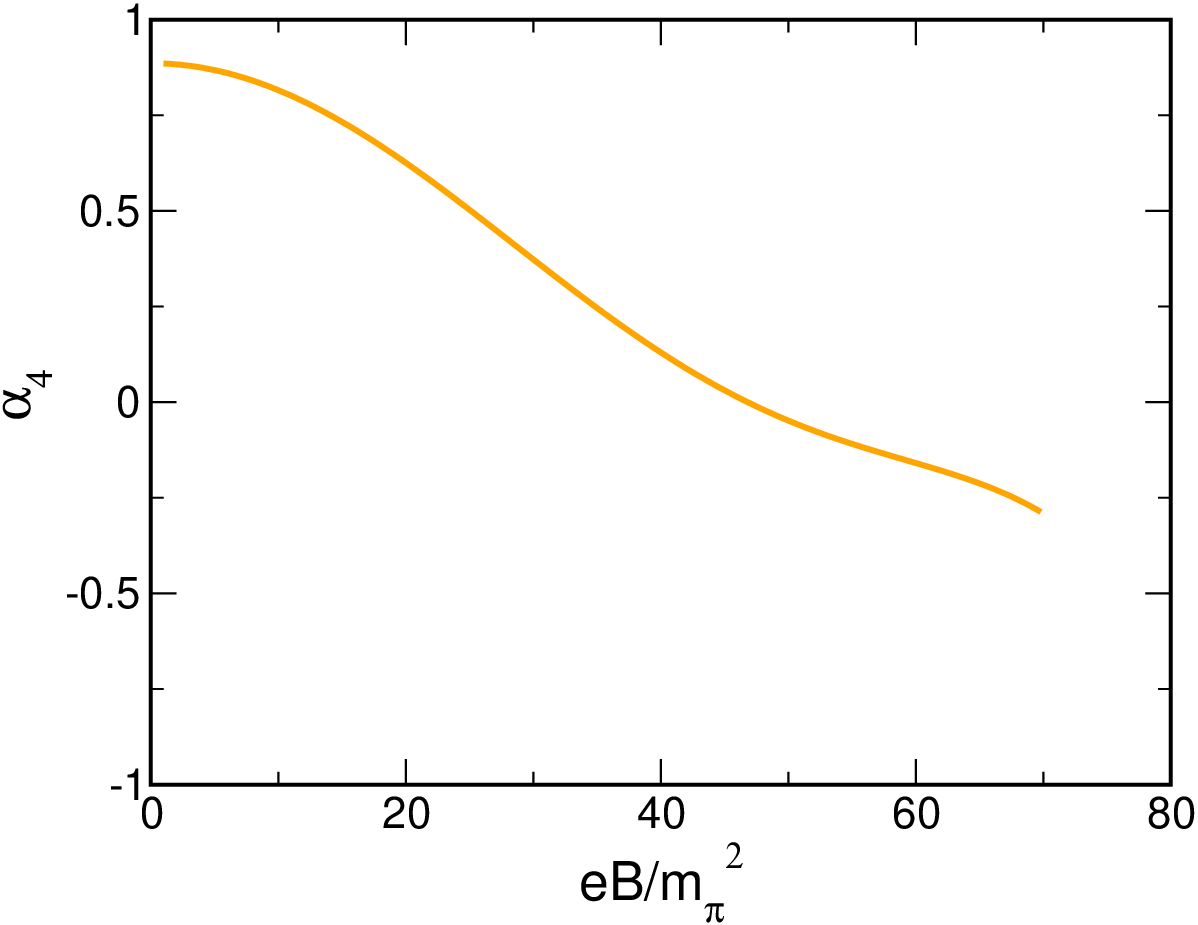}
\caption{\label{Fig:a4}\small Critical temperature ({\em left panel}) 
and $\alpha_4$ at the critical temperature ({\em right panel}) as a function of $eB/m_\pi^2$, 
for the case of the nonrenormalized model. Blue dot corresponds to the critical field, 
$e B_{CP}\approx 47 m_\pi^2$ at which the phase transition changes from a second order to a first order.
The corresponding temperature is $T_{CP} \approx 313$ MeV. The calculation of the critical temperature
in the left panel for $eB > e B_{CP}$ is not reliable, since higher order terms in the GL
expansion should be computed in order to determine the location of the phase transition
in the first order regime.}
\end{center}
\end{figure}

In order to explain how the phase transition changes from second order to first order
in presence of a strong magnetic background,
in Fig.~\ref{Fig:a4} we plot the critical temperature (left panel) 
and $\alpha_4$ at the critical temperature (right panel) as a function of $eB/m_\pi^2$, 
for the case of the nonrenormalized model. We find that even if increasing $eB$ at fixed
temperature results in making $\alpha_4$ more and more positive, in agreement with the
behavior we find for the case of the renormalized model, the value of $\alpha_4$
at the critical temperature is a decreasing function of $eB$. This is due to the behavior
of the critical temperature plotted in the left panel of the figure,
and to the decrease of $\alpha_4$ with the increase of the temperature at fixed $eB$.
Blue dot corresponds to the critical field, 
$e B_{CP}\approx 47 m_\pi^2$ at which the phase transition changes from a second order to a first order.
The corresponding temperature is $T_{CP} \approx 313$ MeV. The calculation of the critical temperature
in the left panel for $eB > e B_{CP}$ is not reliable, since higher order terms in the GL
expansion should be computed in order to determine the location of the phase transition
in the first order regime.

\section{Discussion and Conclusions}
In this article we have computed the effective action of the
chiral condensate around the chiral phase transition, in presence of a strong magnetic background $\bm B$. 
Our goal has been to understand in an analytic fashion the chiral phase transition in presence of
a magnetic field, having in mind the idea of realizing if and how the second order phase transition
at $\bm B=0$ transforms to a first order phase transition at $\bm B \neq 0$. 
We have written the effective action
for the chiral condensate (or equivalently for the dynamical quark mass) around the phase transition 
in a Ginzburg-Landau (GL) form, see Eq.~\eqref{eq:i1}.
For simplicity we have neglected the possibility of a coordinate dependent
condensate, which allows us to neglect the gradient terms in the GL expansion. 
According to the general GL theory of a phase transition, the latter is of the second order
if $\alpha_4>0$, and of first order if $\alpha_4 < 0$;  
the point in the phase diagram with $\alpha_4=0$
is called the critical point. 

In order to map the phase transition lines from the $(\alpha_2,\alpha_4)$
plane to the $(T,\bm B)$ plane we need a specific microscopic model.
To this end we have used firstly the renormalized quark-meson model.
The main scope of our study is to check the existence of a chiral critical point
at $\bm B \neq 0$. The use of the renormalized model allows to make quantitative predictions
which are not affected by an ultraviolet cutoff, which instead affects the predictions of the
nonrenormalizable models. We find that no critical point appears in the phase diagram
in agreement with the lattice simulations,
even if the magnetic field tends to lower the value of the quartic coefficient at the critical
temperature, thus making the phase transition closer to a first order one. 
Our result is in agreement with those of~\cite{Fukushima:2010fe,Skokov:2011ib}
which however are mainly numerical,
while the present work is analytical, and we are able to capture the behavior of the
phase transition in magnetic field in a single GL coefficient.
The power of our computation is even better understood if we restrict ourselves to the LLL approximation, 
which is a good approximation when
$eB >> T^2$, where we can prove analytically that the quartic coefficient is always positive, see Eq.~\eqref{eq:LLL4}. 
This result further excludes the existence
of a critical point at large $\bm B$. We point out that our study
is performed within the one-loop approximation; quantum fluctuations of the meson fields,
which are not taken into account in our study, might modify this conclusion.
However, the numerical computations of~\cite{Skokov:2011ib} seem to
confirm that quantum fluctuations do not affect the qualitative structure of
the phase diagram.

We have also compared the above results with those obtained within the nonrenormalized model,
in which the ultraviolet cutoff is treated as a free parameter and its numerical value
is fixed in order to reproduce few phenomenological quantities. Also in the latter case we have found 
the magnetic field makes the phase transition stronger.
On the other hand, the sign of the quartic coefficient of the GL expansion is affected by
a finite $\bm B$, turning the second order phase transition to a first order one. 
Hence, in the case of the nonrenormalized model, a critical point exists.
Our main conclusion is thus that the existence of the critical point
at finite $\bm B$ is very sensitive to the way the ultraviolet divergences 
of the model are treated. To make this explicit, as well as to summarize the 
main results we obtained, we write here below our final expressions for the
quartic coefficient in the asymptotic limit $eB/T^2 \gg 1$ for the case of renormalized model,
\begin{equation}
\alpha_4 \asymp \frac{3 a_4 N_c}{\pi^2}\sum_f 
\frac{|Q_f e B|}{T^2}~,\label{eq:RF}
\end{equation}
and of the nonrenormalized model,
\begin{equation}
\alpha_4 \asymp \frac{3 a_4 N_c}{\pi^2}\sum_f\frac{|Q_f e B|}{T^2}
\left(
1 - \frac{T^2}{4\Lambda^2 a_4}
\right)~,
\end{equation}
with $a_4 \approx 0.11$. Clearly $\alpha_4$
is positive defined in the case the renormalization procedure is performed;
on the other hand, its sign depends on temperature and in fact it can be negative
if the critical temperature $T_c > 2\Lambda\sqrt{a4}\approx 0.66\Lambda$. 
The latter result however is sensitive to the ultraviolet cutoff, and in fact
the result of the nonrenormalized model tends to that of the renormalized model
if one takes $\Lambda\rightarrow\infty$. However such a limit is not performed
in the numerical computations based on the nonrenormalized model, since in this case
$\Lambda$ is treated as a parameter whose numerical value is fixed once for all
by phenomenological requirements. 

We point out that the use of the nonrenormalized model,
and its cutoff dependent results, is still very interesting for what concerns the applications
to QCD, even if from a pure field theoretical point of view it would be preferrable
to have physical predictions which are not dependent on the
regularization scheme. As we explained in the Introduction, the reason is that the explicit 
UV cutoff appearing in the model results corresponds to a rough modelling
of the QCD asymptotic freedom: the interactions are switched off 
for momenta larger than the UV scale. According to this interpretation of
the model UV scale, our result can be rephrased in the following terms:
the change of the order of the chiral transition at finite temperature and
zero chemical potential, induced by the magnetic field, is mainly connected
to the existence of an intrinsic UV scale, the latter being the remnant
of the QCD asymptotic freedom.

There are several directions which could be followed to extend our work.
One of them is the introduction of baryon chemical
potential $\mu$,
as well as the strange quark with its own chemical potential. In particular, 
since our result suggests that no critical point appears at $\mu = 0$ and
$\bm B\neq 0$, it is important to understand how the critical point 
at $\mu\neq 0$ and $\bm B\neq 0$ develops, following the analysis of~\cite{Andersen:2012bq}. 
Moreover, in our opinion it would be quite important to add a Polyakov loop
background to the model, in order to realize quantitatively if and how the
Polyakov loop affects the chiral phase transition. Qualitatively we do not expect
a dramatic change of the results; on the other hand, a firm statement can be done
only  after the computation is performed. Finally, extending our study
to inhomogeneous condensates following the lines of~\cite{Abuki:2011pf,Flachi:2010yz} 
seems to be an exciting project.

\acknowledgments
M.~R. acknowledges discussions with H. Abuki, G.~Endrody, K. Fukushima, K. Morita,
A. Ohnishi, N.~Su and T. Ueda. We also acknowledge M. Chernodub for correspondence.
The Yukawa Institute for Theoretical Physics and the
Tokyo University of Science are acknowledged for the kind hospitality
during the preparation of the present manuscript.  The work of M.T. is supported by the JSPS Grant-in-Aid
for Scientific Research, Grant No. 24540280.

\end{document}